\documentclass[article,onecolumn,preprintnumbers,amsmath,amssymb,aps,pra,11pt]{revtex4-1}

\usepackage{CJK}
\usepackage{array}
\usepackage{txfonts}
\usepackage{amsmath}
\usepackage{graphicx}
\usepackage{dcolumn}
\usepackage{bm}
\usepackage{ulem}
\usepackage{caption}
\usepackage{hyperref}
\usepackage{setspace}
\hypersetup{
	colorlinks=true,       	 % false: boxed links; true: colored links
	linkcolor=blue,          % color of internal links
	citecolor=blue,           % color of links to bibliography
	urlcolor=blue,            % color of external links
}

\def\ket#1{\left\lvert {#1} \right\rangle}

\def\ketbra#1#2 {\left\langle {#1} | {#2} \right\rangle}  

\begin{document}

\begin{CJK*}{GB}{gbsn}

\title{	Manipulating spatial structure of high-order quantum coherence with entangled photons}
	
\author{Shuang-Yin Huang,$^{1,2}$ Jing Gao,$^{1,2}$ Zhi-Cheng Ren,$^{1,2}$ Zi-Mo Cheng,$^{1,2}$ Wen-Zheng Zhu,$^{1,2}$ Shu-Tian Xue,$^{1,2}$ Yan-Chao Lou,$^{1,2}$ Zhi-Feng Liu,$^{1,2}$ Chao Chen,$^{1,2}$ \\ Fei Zhu,$^{3}$ Li-Ping Yang,$^{4}$ Xi-Lin Wang,$^{1,2,5,6}$ and Hui-Tian Wang$^{1,2,7}$}

\affiliation{ \quad \\
$^{1}$National Laboratory of Solid State Microstructures and School of Physics, Nanjing University, Nanjing 210093, China \\
$^{2}$Collaborative Innovation Center of Advanced Microstructures, Nanjing University, Nanjing 210093, China \\
$^{3}$Intelligent Scientific Systems Co. Limited, Beijing 102208, China \\
$^{4}$Center for Quantum Sciences and School of Physics, Northeast Normal University,\\ Changchun 130024, China \\
$^{5}$Hefei National Laboratory, Hefei 230088, China \\
$^{6}$Synergetic Innovation Center of Quantum Information and Quantum Physics, University of Science and Technology of China, Hefei 230026, China \\
$^{7}$Collaborative Innovation Center of Extreme Optics, Shanxi University, Taiyuan 030006, China}

\date{\today}

\begin{abstract}

\begin{spacing}{1.5}

High-order quantum coherence reveals the statistical correlation of quantum particles. Manipulation of quantum coherence of light in temporal domain enables to produce single-photon source, which has become one of the most important quantum resources. High-order quantum coherence in spatial domain plays a crucial role in a variety of applications, such as quantum imaging, holography and microscopy. However, the active control of high-order spatial quantum coherence remains a challenging task. Here we predict theoretically and demonstrate experimentally the first active manipulation of high-order spatial quantum coherence by mapping the entanglement of spatially structured photons. Our results not only enable to inject new strength into current applications, but also provide new possibilities towards more wide applications of high-order quantum coherence. 
\end{spacing}
\end{abstract}

\maketitle

\end{CJK*}

%\centerline 

%\renewcommand {\thefigure}{S\arabic{figure}}

The high-order coherence of light reveals statistical correlation
beyond the first-order coherence indicated by intensity fringes in traditional interference (for instance, the best known Young's double-slit interference). Study on high-order coherence was pioneered by Hanbury Brown and Twiss (HBT) in a landmark interferometry experiment in 1956~\cite{Brown1956}, in which they observed the bunching of photons in thermal light. This discovery led Glauber to lay the foundation of quantum optics by developing the quantum theory of coherence~\cite{Glauber1963} and found practical application in astronomy to measure the angular diameter of stars~\cite{Brown1956-2}. Since then, the HBT interferometry has been deeply explored with various sources, such as interacting photons in a nonlinear medium~\cite{Bromberg2010} and twisted light~\cite{Loaiza2016}, and has attracted significant interest beyond photons, including electrons (antibunching effect for fermions)~\cite{Henny1999, Oliver1999}, atoms~\cite{Schellekens2005, Ottl2005, Jeltes2007}, molecules~\cite{Rosenberg2022}, matter wave~\cite{Perrin2012} and phonons~\cite{Cayla2020}.  

\begin{figure}
    \centering
    \includegraphics[width=0.8\linewidth]{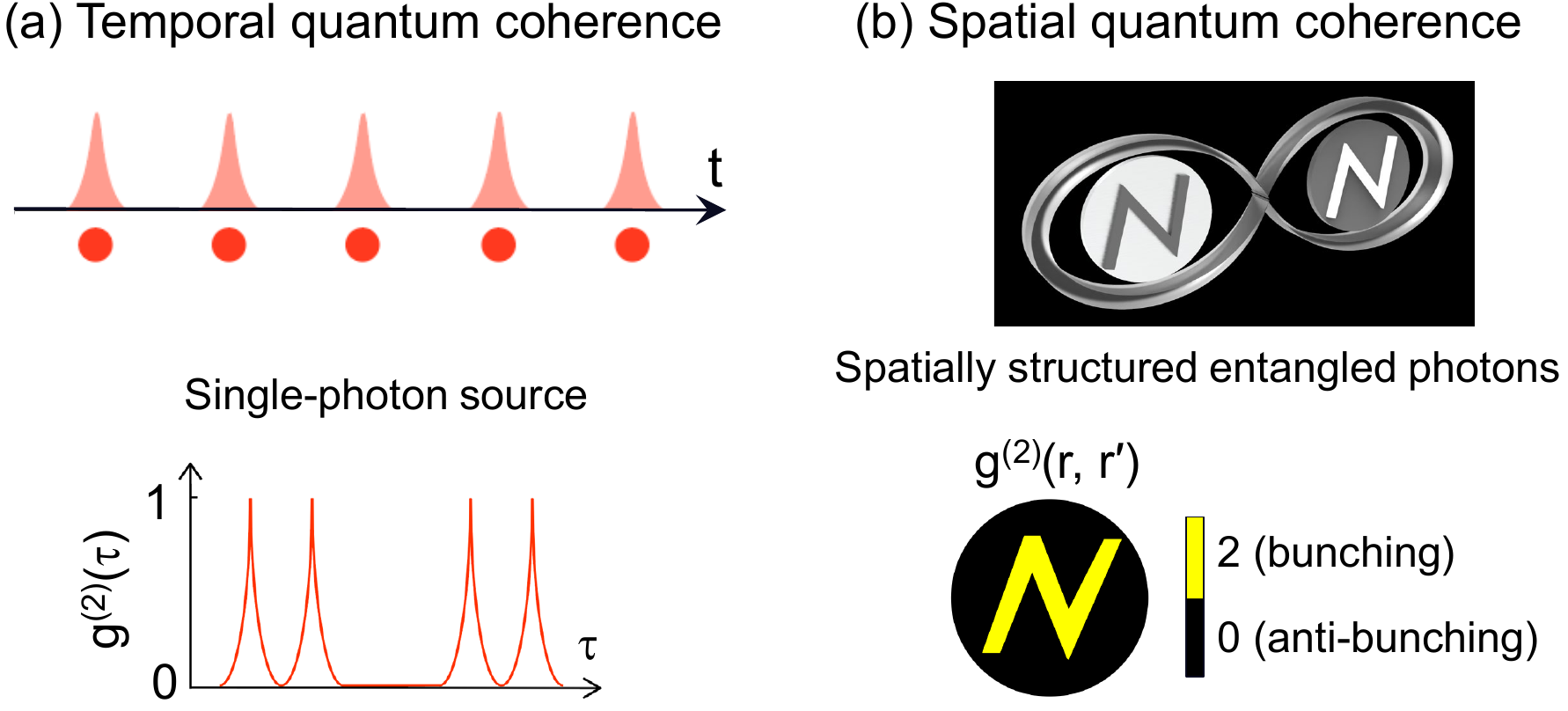}   
    \caption{\label{fig:1}Temporal coherence versus spatial coherence. (a) Traditionally, the second-order temporal coherence function $g^{(2)}(\tau)$ is used to describe the bunching and antibunching statistics of quasi-one-dimensional photon pulses. (b) The second-order spatial coherence function $g^{(2)}(\mathbf{r}, \mathbf{r}')$ characterizes the quantum spatial correlations of spatially structured entangled photons.
    }
\end{figure}

In contrast to the thermal source, the high-order coherence of
quantum light unveils remarkable statistical behaviors among
quantum particles. One of the typical quantum sources is the single-photon source~\cite{He2013, Ding2016, Somaschi2016, Tomm2021}, which shows the antibunching effect in temporal domain and has important applications in quantum information science ranging from quantum foundation~\cite{Tang2012, Wen2023}, quantum communications~\cite{Lu2021}, quantum metrology~\cite{Chu2017, Wang2020} to quantum computation and quantum simulation~\cite{Wang2019, Cogan2023}. The bunching and antibunching behaviors of photons are usually described by the second-order quantum coherence in temporal domain, which is quantified by $g^2(\tau)|_{\tau =0}$, where $\tau = t_2 - t_1$ indicates the arriving time difference between the two detected photons. A perfect single-photon source exhibits a vanishing $g^2(\tau)|_{\tau =0} = 0$, indicating its intrinsic ``single-photon'' nature, as illustrated in Fig.~\ref{fig:1}(a). The antibunching effect, characterized by $g^2(\tau)|_{\tau =0} < 1$, originates from the sub-Poissonian photon statistics~\cite{Short1983} and signifies a genuine quantum phenomenon with no classical counterpart. Therefore, the control of high-order quantum coherence in \textit{temporal domain} has brought significant impact on quantum information science and quantum optics. Consequently, the manipulation of this quantum feature in \textit{spatial domain} would be expected to introduce new quantum effects and applications.

From the perspective of spatial domain, the existing research on the manipulation of high-order quantum coherence was mainly focused on the \textit{uniform} spatial modes. By establishing entanglement in spatial degree of freedom (DOF), for example, orbital angular momentum (OAM) arising from a helical phase in the form of $\exp (j m \varphi)$ (where $\varphi$ is the azimuthal angle in the polar coordinates and the integer index $m$ indicates the topological charge), one can construct spatially structured entangled photons~\cite{Mair2001, Kovlakov2017, Defienne2018, Erhard2020, Forbes2021}. The quantum feature from the nonuniform spatial structure can be utilized to produce various types of entanglement, including two-photon~\cite{Dada2011} and multi-photon~\cite{Zhang2016, Hiesmayr2016, Malik2016, Zhang2017, Erhard2018} high-dimensional entanglement, high-angular-momenta entanglement~\cite{Fickler2012, Hiekkamaki2021}, hyperentanglement in multiple DOFs~\cite{Barreiro2005, Zhao2019}. These novel quantum entanglements have found many important applications in various quantum information operations and protocols, such as  full Bell state measurement~\cite{Barreiro2008}, alignment-free quantum communication~\cite{D'Ambrosio2012}, direct characterization of quantum dynamics~\cite{Graham2013}, quantum teleportation of multiple DOFs of a single photon~\cite{Wang2015}, and direct measurement of particle exchange phase~\cite{Liu2022}. Exploring the spatial nonuniformity in structured entangled photons has emerged as an appealing and valuable realm. Moreover, the spatial nonuniformity in high-order quantum coherence holds great potential for intriguing effects and significant applications. However, on-demand manipulation of achieving high-order quantum coherence remains challenging. 

In this letter, we theoretically propose and experimentally demonstrate the active control of the nonuniform high-order quantum coherence of spatially structured entangled photons, as shown in Fig.~\ref{fig:1}(b). In theory, we develop an approach that maps the entanglement of spatially structured photons to the nonuniform high-order quantum coherence, which is applicable for both two and more entangled photons. In experiment, we successfully verified our approach by measuring the second-order spatial quantum coherence $g^{(2)}(\mathbf{r},\mathbf{r}')$ of spatially entangled two photons at respective positions $\mathbf{r}$ and $\mathbf{r}'$. The key to our experiment is to transfer the polarization entanglement of two spatially uncorrelated photons to the spatially structured entanglement. Our results have the potential to pave the way for new possibilities in quantum information processing and enhance the control over photon-based systems.

\begin{figure}
    \centering \includegraphics[width=0.6\linewidth]{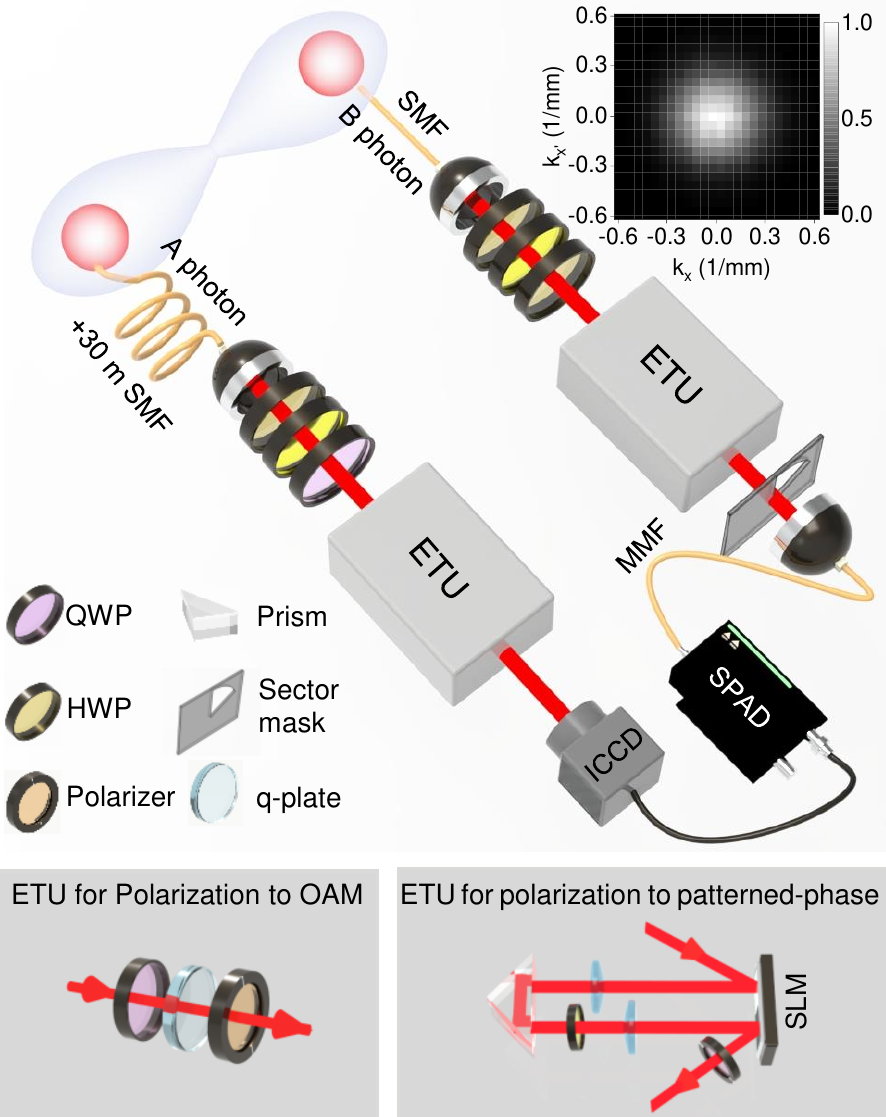}   
    \caption{\label{fig:2}Experimental setup of manipulating nonuniform high-order quantum coherence of spatially structured entangled photons. Using single-mode fibers (SMFs), we eliminate the momentum correlation of the photon pair, as depicted in the inset. The two-photon polarization entanglement is converted into either OAM entanglement or patterned-phase entanglement via the entanglement transfer units (ETUs), which behaves as a polarization-dependent mode switching device. The idler photon in path-B is filtered by a $\pi$/4-sector mask  and collected by a multimode fiber (MMF) connected to a single-photon avalanche diode (SPAD). The electrical signal of SPAD is used to trigger the ICCD camera realizing the coincidence imaging of the signal photon in path-A. 
    }
\end{figure}

\textit{Manipulation of nonuniform quantum coherence}.---The essence of our quantum coherence control method lies in the conversion of global polarization entanglement into nonuniform spatial correlation. The resulting patterned-phase entangled two-photon states are described by a wave-packet function (WPF) of the form~\cite{Cui2023}
\begin{equation}
\Psi (\mathbf{r},\mathbf{r}') = \frac{1}{\sqrt{2}}\eta_A(\mathbf{r})\eta_B(\mathbf{r}')\left\{e^{i[\Phi_A(\mathbf{r})-\Phi_B(\mathbf{r}')]} \pm {\rm c.c.}\right\}, \label{eq:WPS} 
\end{equation}
where the normalized function $\eta_{i} (\mathbf{r})$ $i = \{A, B\}$ represents the probability amplitude distribution of the paraxial photons in path-$i$. By engineering the patterns in phases $\Phi_A(\mathbf{r})$ and  $\Phi_B(\mathbf{r})$, we can precisely control the spatial structure of second-order coherence function of a photon pair (see Supplemental Material for details~\cite{SuppMat})
\begin{equation}
 g^{(2)}(\mathbf{r},\mathbf{r}') = \frac{G^{(2)}(\mathbf{r},\mathbf{r}')}{n_A(\mathbf{r})n_B (\mathbf{r}')}=1 \pm \cos [2\Phi_A(\mathbf{r}_A)-2\Phi_B(\mathbf{r}'_B)].\label{eq:g2}
\end{equation}
Here, $G^{(2)}(\mathbf{r},\mathbf{r}')$ represents the second-order correlation function~\cite{Glauber1963}. We note that the photon number density distribution of each individual photon remains almost unchanged despite the presence of the nonuniform spatial entanglement, i.e., $n_A (\mathbf{r})\approx |\eta_A (\mathbf{r})|^2$ and $n_B (\mathbf{r}')\approx |\eta_B (\mathbf{r}')|^2$ in the paraxial case.

\begin{figure}
    \centering  \includegraphics[width=0.8\linewidth]{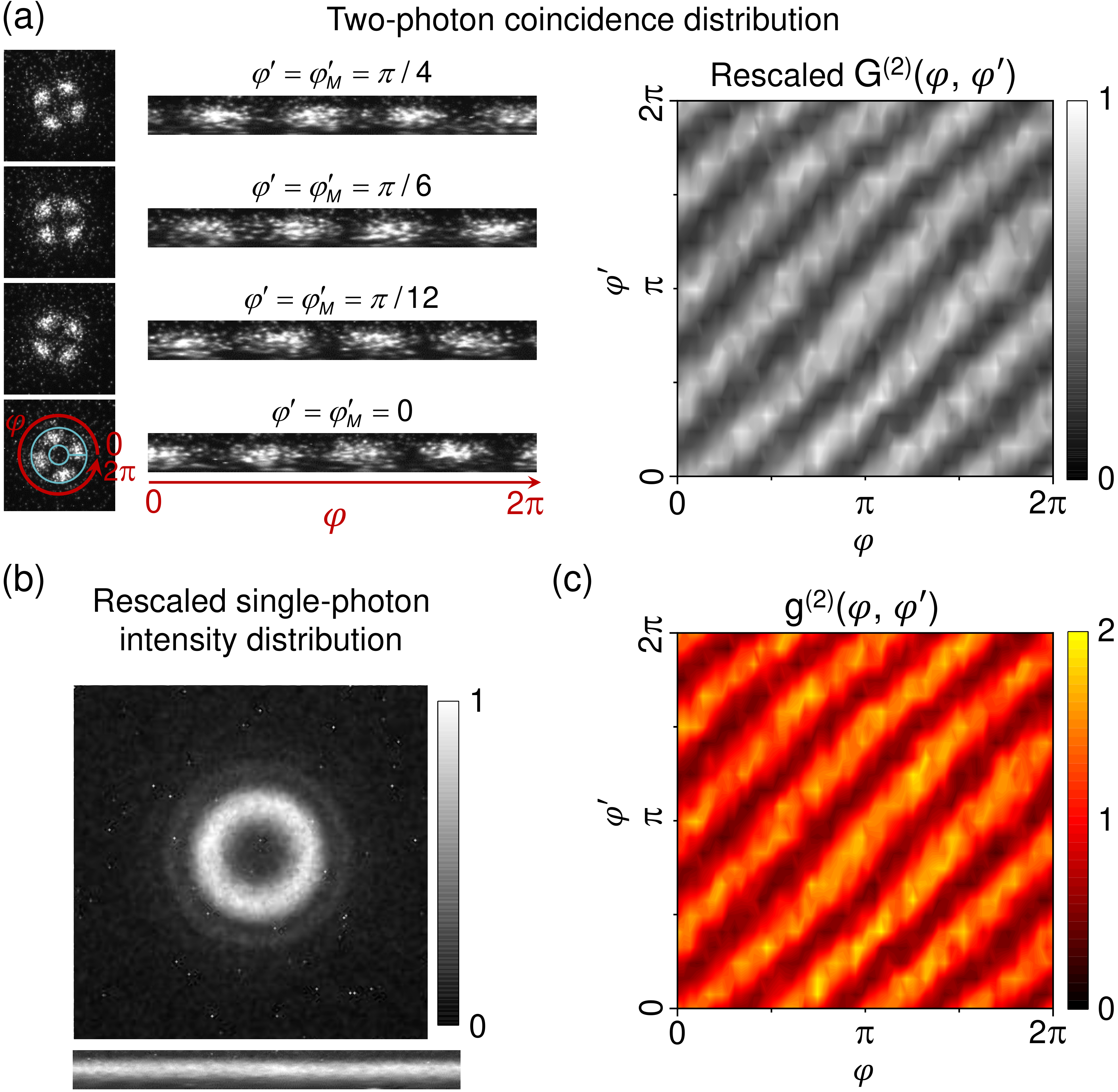}   
    \caption{\label{fig:3} The contrast between photon density distribution, two-photon correlation $G^{(2)}(\mathbf{r},\mathbf{r}')$, and coherence function $g^{(2)} (\mathbf{r},\mathbf{r}')$ for helical phases $\Phi_A (\mathbf{r})=2\varphi$ and $\Phi_B (\mathbf{r}')=2\varphi'$. (a) The left column shows the coincidence patterns captured by the ICCD camera at path-A triggered by the photon in path-B with the sector mask at four different orientation angles ($\varphi'=\pi/4,\pi/6,\pi/12,0$). The middle column displays the corresponding unfolded patterns of the left column. The right column presents the rescaled two-photon coincidence pattern $G^{(2)} (\varphi,\varphi')$. (b) The rescaled single-photon intensity distribution directly captured by the ICCD exhibits a doughnut structure and the corresponding unfolded pattern is displayed below. (c) The obtained second-order coherence function $g^{(2)}(\varphi,\varphi')$.
 }
\end{figure}

\textit{Experimental realization}.---In our experiment, as depicted in Fig.~\ref{fig:2}, a mode-locked laser with a repetition rate of 80 MHz and a wavelength of 390 nm pumps a sandwich-like type-II crystal composed of a 4-mm-long BBO+HWP+BBO setup~\cite{Wang2016}. By using the degenerate and non-collinear spontaneous parametric down-conversion (SPDC) process, we generate the polarization-entangled photon pairs in the form of $(|H_AV_B\rangle\pm|V_AH_B\rangle)/\sqrt{2}$, where $H$ and $V$ indicate horizontally and vertically polarized states, respectively. We utilize the single-mode fibers (SMFs) and 3-nm narrow-band filters to eliminate the momentum and frequency correlations of the photon pair. The obtained joint momentum spectrum exhibits a Gaussian distribution~\cite{Walborn2010}, as shown in the inset at the top-right corner, implying the elimination of momentum correlation. We achieve the polarization entanglement fidelity of $\sim$$0.982$ after the two photons passing through the SMFs.

We utilize the entanglement transfer units (ETUs) composed of polarization-dependent mode switching devices to convert the polarization entanglement into the patterned-phase entanglement. The phases $\Phi_i (\mathbf{r})$ ($i={A,B}$) and $-\Phi_i (\mathbf{r})$ are applied to the $H$- and $V$-polarized photons, respectively, using q-plates~\cite{Marrucci2006} or spatial light modulators (SLMs). Next, we select the $H$-polarized photons described by the WPF in Eq.~(\ref{eq:WPS}) for coincidence, using the polarization beam splitters (PBSs).

In the coincidence measurement, the idler photons in path-B are filtered by a mask and collected by a multimode fiber (MMF), which is connected to a single-photon avalanche diode (SPAD). The SPAD detector serves as a trigger for a pixel-intensified charge-coupled device (ICCD) camera (2DSPC-GS-F), and enables to capture the image of the spatial coincidence distribution of the signal photons in path-A, i.e., the $G^{(2)}(\mathbf{r},\mathbf{r}')$ function with fixed $\mathbf{r}'$. To ensure that the detected photons at the heralding detector and the camera originate from the same correlated photon pair, an additional ($\sim$30 meter) SMF is used to compensate for the electrical delay (150~ns) that occurs during the ICCD triggering process. In experiment, we insert a $\pi/4$-sector mask (instead of a very narrow sector mask) in path-B to obtain the stronger coincidence signal, in which the orientation angle of the angular bisector of the sector mask is defined by $\varphi'_M$. 

To demonstrate the nonuniform coherence of OAM entanglement, we apply helical phases $\Phi_A (\mathbf{r})=m_A\varphi$ and $\Phi_B (\mathbf{r}')=m_B\varphi'$ to the two photons via q-plates. Here, $\varphi$ and $\varphi'$ are the azimuthal angles of the two cylindrical coordinate systems associated with the two paraxial photons. The resulting OAM-entangled states are given by $(\ket{m_A, -m_B} \pm \ket{-m_A, m_B})/\sqrt{2}$. According to our theory, the second-order correlation function will exhibit the azimuthal nonuniformity, i.e., $g^{(2)}(\mathbf{r},\mathbf{r}')=g^{(2)}(\varphi,\varphi')=1\pm\cos[2(m_A\varphi-m_B\varphi')]$.

Figure~\ref{fig:3} illustrates the experimental results for the OAM-entangled state $(\ket{2_A,-2_B} + \ket{-2_A,2_B})/\sqrt{2}$, i.e., $m_A=m_B=2$. In the left column of Fig.~\ref{fig:3}(a), we present the spatial coincidence distribution captured by the ICCD camera. Notably, the coincidence image exhibits a $C_4$ symmetry for any fixed $\varphi' = \varphi'_M$, and the patterns rotate with the orientation angle $\varphi'_M$ of the $\pi/4$-sector mask in path-B. The middle column of Fig.~\ref{fig:3}(a) displays the corresponding unfolded patterns of the left column~\cite{MoreauSA2019}, in which each row corresponds to four different orientation angles of $\varphi'_M$. Furthermore, in the right column of Fig.~\ref{fig:3}(a), we present the rescaled second-order correlation function $G^{(2)}(\varphi,\varphi')$ as a function of the angles $\varphi$ and $\varphi'$. 

\begin{figure}
    \centering \includegraphics[width=0.8\linewidth]{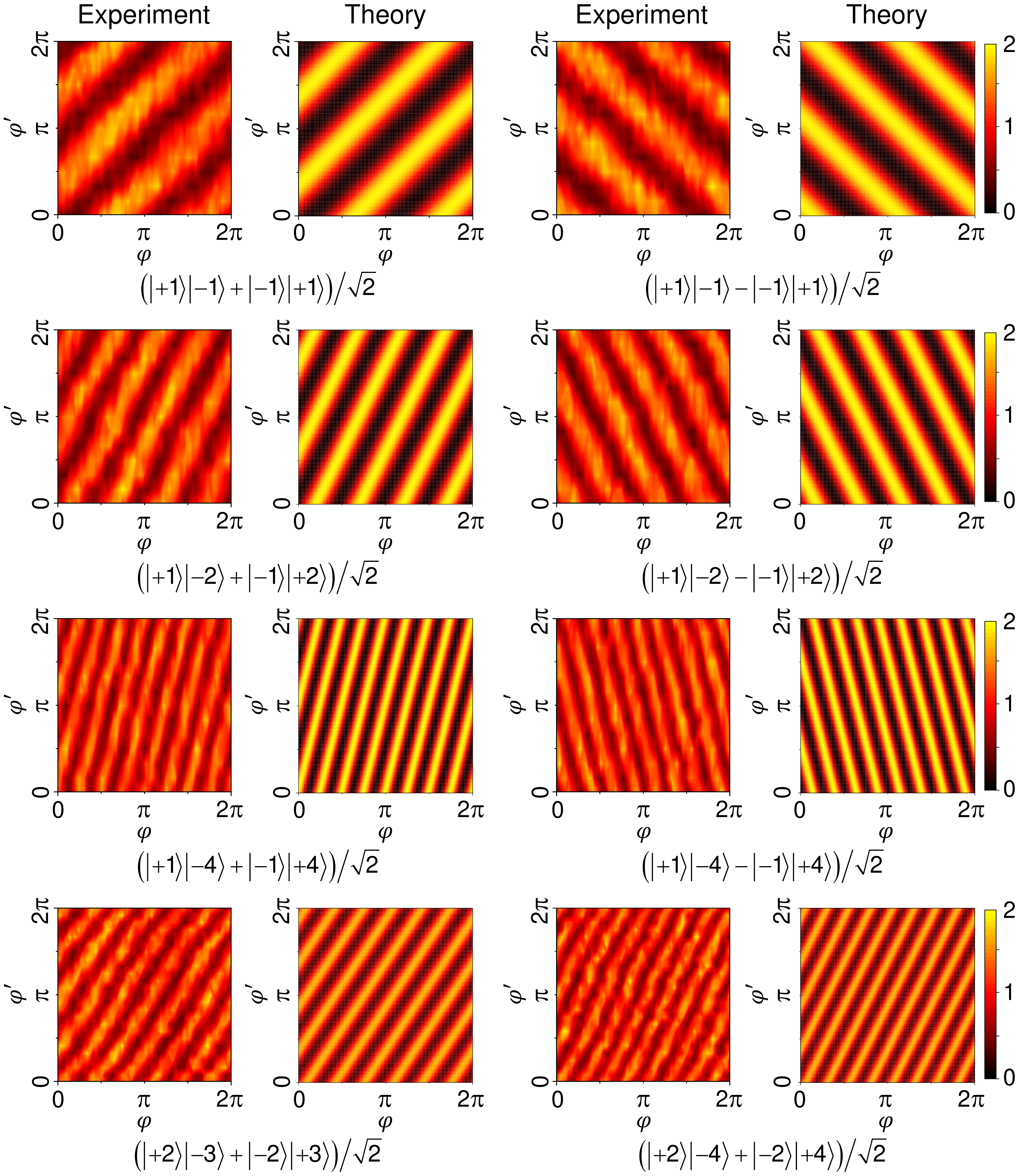}   
    \caption{\label{fig:4} The second-order coherence function $g^{(2)} (\varphi,\varphi')$ for different OAM entangled states $(\ket{m_A,-m_B} \pm \ket{-m_A,m_B})/\sqrt{2}$.
    }
\end{figure}

\begin{figure}
    \centering \includegraphics[width=0.6\linewidth]{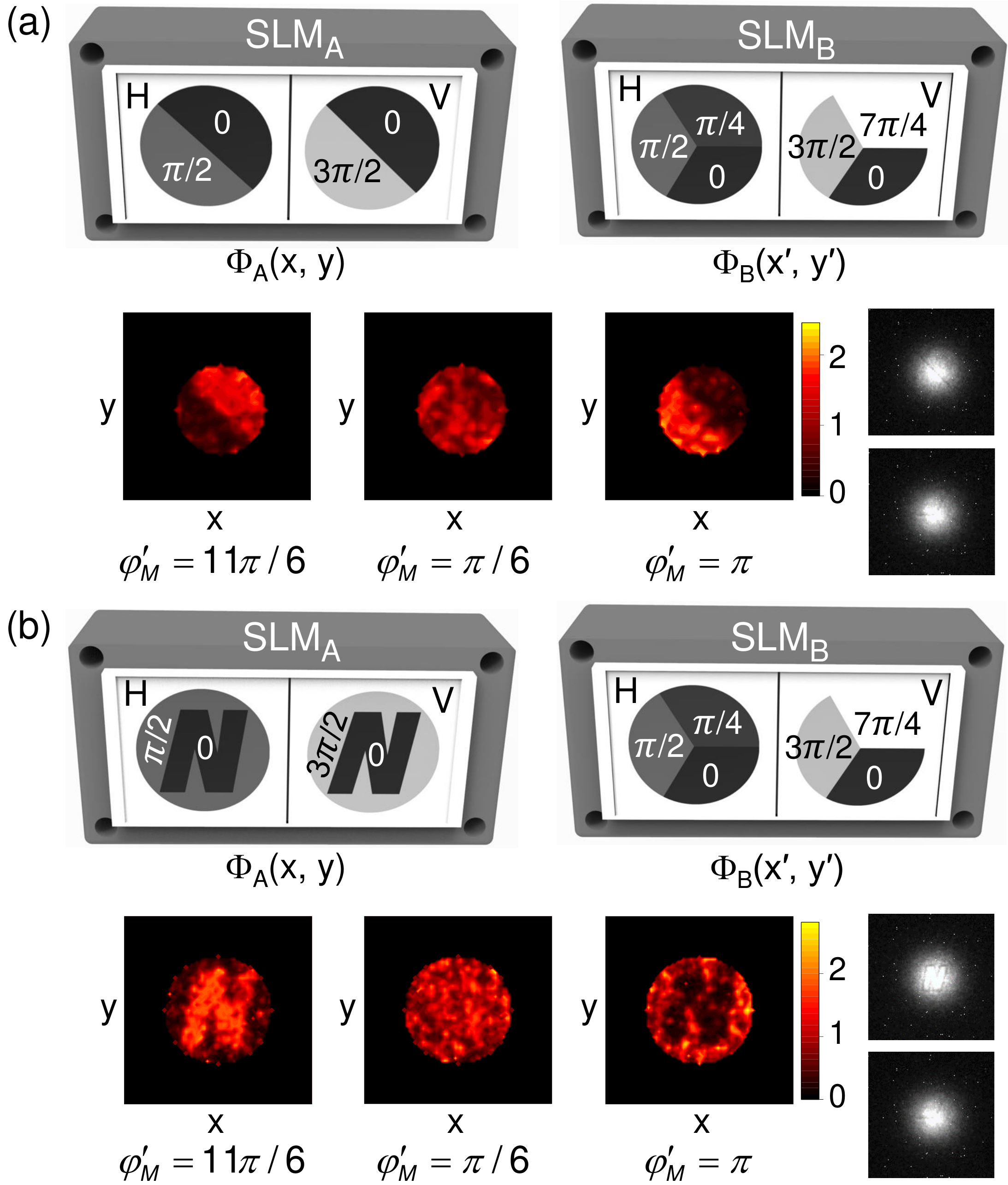}   
    \caption{ \label{fig:5} Imposing arbitrary patterned-phase in the coherence function of an entangled photon pair. (a) The $g^{(2)}$ function for sector-shaped patterned-phases. The first row illustrates the imposed patterned-phases $\Phi_i (\varphi)$ and $- \Phi_i (\varphi)$ ($i=\{A,B\}$) for $H$- and $V$-polarized photons, respectively. The second row presents the measured $g^{(2)}$ functions for different scanning phase $\Phi_B=\{0,\pi/4,\pi/2\}$ in the first three figures. Neither the signal photon nor the idler photon exhibits any sector structure in their photon-number density. (b) The $g^{(2)}$-function of photon pairs subjected to arbitrary-pattern entanglement.
} 
\end{figure}

As predicted by our theory, the $C_4$-symmetric pattern does not appear in the photon-number density distribution. Without coincidence, only the doughnut structure of the OAM photon is observed, as shown in Fig~\ref{fig:3}(b). To extract the $g^{(2)}$ function in our experiment, we employ the following formula
\begin{equation}
g^{(2)}(\varphi,\varphi')=\frac{P_{AB} (\varphi, \varphi')} {P_A (\varphi)},
\end{equation}
where $P_{AB}$ represents the normalized probability distribution of the path-A photons in coincidence measurement, while $P_{A}$ refers to the normalized probability distribution of the path-A photon in the absence of coincidence. The measured coherence function $g^{(2)}(\varphi,\varphi')$, as shown in Fig.~\ref{fig:3}(c), is in good agreement with the theoretical prediction given in Eq.~(\ref{eq:g2}) and exhibits the similar fringes to the $G^{(2)} (\varphi,\varphi')$ function in panel (a). 

Our approach can be applied to cases when the signal and idler photons have different OAM quantum numbers. To demonstrate this, we present a comparison between experimental results and theoretical simulations for various cases in Fig.~\ref{fig:4}. The quantity of fringes is determined only by the OAM quantum number ($m_A$) of the signal photons and is independent of $m_B$ of the idler photons. The fringes with positive slope are observed for the state of $(\ket{m_A, -m_B} + \ket{-m_A, m_B})/\sqrt{2}$, while the fringes with negative slope are observed for the state of $(\ket{m_A, -m_B} - \ket{-m_A, m_B})/\sqrt{2}$. Obviously, the measured $g^{(2)}$ functions match the theoretical results very well.

\textit{Tailoring quantum coherence}.---Our approach offers the precise control over the spatial structure of quantum coherence in photon pairs. We impose the conjugate phases to the orthogonally polarized photons. With the aid of the ETU for polarization to patterned-phase, we are able to tailor arbitrary structures in the $g^{(2)}$ function of entangled photon pairs.

In Fig.~\ref{fig:5}(a), the two areas of SLM$_A$ in path-A are loaded a binary phase ($0,\pi/2$) for the $H$-polarized photons and its conjugate phase ($0, -\pi/2$) [which is equivalent to ($0,3\pi/2$)] for the $V$-polarized photons. While the two areas of SLM$_B$ in path-B are loaded a ternary phase ($0,\pi/4,\pi/2$) for the $H$-polarized photons and its conjugate phase ($0,-\pi/4,-\pi/2$) [which is equivalent to ($0,7\pi/4,3\pi/2$)] for the $V$-polarized photons, respectively. By rotating the sector mask, we can change the phase $\Phi_B|_{\varphi' = \varphi'_M = (11\pi/6, \pi/6, \pi)} = (0, \pi/4, \pi/2)$ of the idler photons in path-B, we obtained three different distributions of the $g^{(2)}$ function as shown by the three images in the second row. The observed three different values of the coherence function $g^{(2)}(\varphi,\varphi') = 0, 1, 2$ represent the destructive, uniform and constructive distributions according to Eq.~(\ref{eq:g2}). Neither the signal photons nor the idler photons exhibits any sector structure in their photon-number density distributions, as shown in the inset at the bottom right corner.

Similarly, we can apply the binary phases with arbitrary patterns to the signal photons in path-A, such as the letter ``N''. By using the same procedure as Fig.~\ref{fig:5}(a), we observe the letter ``N'', its complementary pattern, and a uniform spot in the $g^{(2)}(\varphi,\varphi')$ function, as depicted in Fig.~\ref{fig:5}(b). 

In conclusion, we have for the first time demonstrated the active manipulation of nonuniform high-order quantum coherence in both theory and experiment. The manipulation is realized by mapping the entanglement of spatially structured entangled photons to the high-order spatial quantum coherence. The on-demand spatially structured entangled photons are generated with polarization-entangled and spatial-uncorrelated photons through entanglement transferring from polarization to OAM (using q-plates) or a patterned phase (using SLMs). The experimental results clearly verified the feasibility of our approach. High-order quantum coherence in spatial domain has played an key role in many quantum applications ranging from to imaging~\cite{Edgar2012, Defienne2018-2, Moreau2019, Defienne2021,CuiYi2023}, holography~\cite{Chrapkiewicz2016, Defienne2021-2} to microscopy~\cite{Tenne2019, Varnavski2020, Lubin2022, Ndagano2022}. Active control of spatial high-order quantum coherence would not only enable to tap more potential in these applications but also open new opportunities for quantum information and quantum optics based on high-order coherence. Furthermore, our approach can be expanded to higher-order quantum coherence control, such as the fourth-order coherence function of four-photon pulses.

%\bibliography{main}
%%%%%%%%%%%%%%%%%%%%%%%%%%%%%%%%%%%%%%%%%%%%%%%%%%%%%%%%%%%%%%%%%%%
\begin{acknowledgments}
This work was supported by the National Key R\&D Program of China (Grants No. 2019YFA0308700 and No. 2020YFA0309500); National Natural Science Foundation of China (Grants No. 12234009, No.12275048 and No. 12274215); Innovation Program for Quantum Science and Technology (Grant No. 2021ZD0301400); Program for Innovative Talents and Entrepreneurs in Jiangsu; Natural Science Foundation of Jiangsu Province (Grant No. BK20220759); Key R\&D Program of Guangdong Province (Grant No. 2020B0303010001).
\end{acknowledgments}

%%%%%%%%%%%%%%%%%%%%%%%%%%%%%%%%%%%

\end{document}